# Cross-study Reliability of the Open Card Sorting Method

**Christos Katsanos**
Department of Informatics, Aristotle University of Thessaloniki, Thessaloniki, Greece
ckatsanos@csd.auth.gr

**Nikolaos Avouris**
Department of Electrical and Computer Engineering, University of Patras, Patras, Greece
avouris@upatras.gr

**Ioannis Stamelos**
Department of Informatics, Aristotle University of Thessaloniki, Thessaloniki, Greece
stamelos@csd.auth.gr

**Nikolaos Tselios**
Educational Sciences and Early Childhood Education Department, University of Patras, Patras, Greece
nitse@ece.upatras.gr

**Stavros Demetriadis**
Department of Informatics, Aristotle University of Thessaloniki, Thessaloniki, Greece
sdemetri@csd.auth.gr

**Lefteris Angelis**
Department of Informatics, Aristotle University of Thessaloniki, Thessaloniki, Greece
lef@csd.auth.gr

## ABSTRACT

Information architecture forms the foundation of users' navigation experience. Open card sorting is a widely-used method to create information architectures based on users' groupings of the content. However, little is known about the method's cross-study reliability: Does it produce consistent content groupings for similar profile participants involved in different card sort studies?

**KEYWORDS**

Card sorting; Information architecture; Website structure; Method evaluation

This paper presents an empirical evaluation of the method's cross-study reliability. Six card sorts involving 140 participants were conducted: three open sorts for a travel website, and three for an eshop. Results showed that participants provided highly similar card sorting data for the same content. A rather high agreement of the produced navigation schemes was also found. These findings provide support for the cross-study reliability of the open card sorting method.

**INTRODUCTION**

Users come to a website with some information needs and expectations about where to look for information. Information architecture (IA) represents the underlying structures that give shape and meaning to the content and functionality on a website [3]. User-centered IA aims to increase findability of information [7, 8] and, in turn enhance the user experience.

The most widely adopted method to create user-centered IAs is card sorting [8, 9, 11, 21]. Card sorting "can provide insight into users' mental models, illuminating the way that they often tacitly group, sort and label tasks and content within their own heads" [8].

In card sorting, representative users are given a stack of index cards. Each card contains one word or phrase representing the information or services provided on webpages. Participants are asked to group the cards in stacks that make sense to them and subsequently name the resulting groups. There are two primary alternatives, open and closed card sorting. The difference lies in the existence or not of a pre-established initial set of groups. The method can be applied in a typical in-person session or by using suitable tools designed to moderate the process remotely [11, 15]. A typical step-by-step roadmap to effectively apply the open card sorting method is described in [15].

A large body of literature confirms the usefulness of the open card sorting method and investigates various related factors, such as the number of minimum required users per card sort, how to address the complexity of sorting many cards, and cultural issues [1, 10, 13, 18, 21]. If there are distinct user groups then one should involve enough participants for each group, analyze the data separately and compare results [17]. This is similar to running multiple card sort studies due to the potential existence of vastly different mental models among participants in distinct groups.

However, how consistent are the groupings provided by similar profile participants? Is it enough to run one card sort study? Despite the popularity of the card sorting method, its cross-study reliability has not been examined. This is the aim of this paper. To this end, six studies were conducted: three open sorts for a travel website and three for an eshop.

This paper addresses the following research questions:
- RQ1: Does the card sorting method produce consistent content groupings for similar profile participants engaged in different card sort studies?
- RQ2: Does quantitative analysis (clustering) of card sorting groupings provide consistent findings for such different card sort studies?
- RQ3: Is there a high level of agreement of navigation schemes produced by such different card sort studies?

Table 1: Studies Overview

| No | Reference Code | N (All) | N (Male) | N (Female) |
|---|---|---|---|---|
| 1 | Travel1 | 34 | 25 | 9 |
| 2 | Travel2 | 26 | 14 | 12 |
| 3 | Travel3 | 22 | 2 | 20 |
| 4 | Eshop1 | 22 | 6 | 16 |
| 5 | Eshop2 | 16 | 4 | 12 |
| 6 | Eshop3 | 20 | 17 | 3 |
| Sum1 | TravelAll | 82 | 41 | 41 |
| Sum2 | EshopAll | 58 | 27 | 31 |
| Sum3 | StudiesAll | 140 | 68 | 72 |

## METHODOLOGY

All in all, six studies involving 140 users were conducted (Table 1). All participants shared a tech-savvy profile being either technology students or practitioners. They were recruited through university courses, seminars and events, such as World Usability Day.

The first set of three open sorts, hereafter Travel1, Travel2 and Travel3, dealt with the design of a travel website. Eighty-two users, 41 female, aged from 21 to 55 (M=25.9, SD=2.5) were asked to sort 38 content items, such as "Travel Planner", "Cuisine", and "Accommodation Options". The second set of three open sorts, hereafter Eshop1, Eshop2 and Eshop3, dealt with the IA redesign of an existing popular eshop selling technology products, and office furniture and consumables. Fifty-eight participants, 31 female, aged from 21 to 49 (M=24.7, SD=3.8) were asked to sort 55 content items, such as "Laptops", "Televisions", "Chairs", "Digital frames", and "Pens & Pencils".

Both study domains are general-purpose and do not require any specialized knowledge. Within each set of studies, users were provided with the same titles and descriptions for the webpages. The number of participants ranged from 16 to 34 on a per study basis. Research suggests a minimum of 15 users to obtain robust data from open card sorts [10, 18], thus our studies had adequate sample size.

In each study, participants first provided their informed consent, then completed a short questionnaire with demographic-related questions (gender, age) and finally performed a typical open card sort. Google Forms was used to create and distribute the questionnaire. Either USort [2] or OptimalSort was used by participants to facilitate the card sort.

Analysis of the collected data was supported by EZCalc [2], Excel 2010 and SPSS Statistics v20.0. The assumption of normality was investigated using Shapiro-Wilk tests and Q-Q plots. If it was violated, then the non-parametric equivalent statistical test was employed.

## RESULTS AND DISCUSSION

### RQ1: Groupings from Different Card Sorting Studies

The main quantitative data from a card sorting study is a similarity score per pair of content items [15]. If all participants grouped two content items together then these items have 100% similarity, whereas if no users placed them together then they have 0% similarity. These similarity scores are typically grouped in an N×N distance matrix where N is the number of cards and each cell contains a dissimilarity score.

The question is how to compare the distance matrices produced for the same content items by participants in different open sort exercises. One approach would be to consider all of the distances of cards as pairs and use correlation analysis. However, this approach would violate the assumption of independence of observations since, for example, the distance between card 1 and 2 is not independent of 1 and 3. There is research [16] suggesting that the correlation of two distance matrices should be tested with the Mantel test [6]. This permutation test overcomes the assumption of independence as it is designed for the null hypothesis that two distance matrices of the same cases are uncorrelated. In the following, we report both approaches.

**Table 2: Spearman Correlations for the Distance Matrices of the Travel Card Sorts**

|         | Travel1  | Travel2  | Travel3 |
|---------|----------|----------|---------|
| Travel1 | 1        |          |         |
| Travel2 | 0.727**  | 1        |         |
| Travel3 | 0.419**  | 0.432**  | 1       |

**Correlation is significant at 0.01 level (2-tailed)

**Table 3: Mantel Tests (10000 permutations) for the Distance Matrices of the Travel Card Sorts**

|         | Travel1   | Travel2   | Travel3 |
|---------|-----------|-----------|---------|
| Travel1 | 1         |           |         |
| Travel2 | 0.894***  | 1         |         |
| Travel3 | 0.609***  | 0.633***  | 1       |

***Correlation is significant at 0.001 level (2-tailed)

**Table 4: Spearman Correlations for the Distance Matrices of the Eshop Card Sorts**

|        | Eshop1   | Eshop2   | Eshop3 |
|--------|----------|----------|--------|
| Eshop1 | 1        |          |        |
| Eshop2 | 0.650**  | 1        |        |
| Eshop3 | 0.851**  | 0.844**  | 1      |

**Correlation is significant at 0.01 level (2-tailed)

**Table 5: Mantel Tests (10000 permutations) for the Distance Matrices of the Eshop Card Sorts**

|        | Eshop1    | Eshop2    | Eshop3 |
|--------|-----------|-----------|--------|
| Eshop1 | 1         |           |        |
| Eshop2 | 0.751***  | 1         |        |
| Eshop3 | 0.930***  | 0.873***  | 1      |

***Correlation is significant at 0.001 level (2-tailed)

**Table 6: Base-clusters Comparisons**

|                    | Separation | Similarity |
|--------------------|------------|------------|
| Travel1 vs Travel2 | 4.7%       | 95.3%      |
| Travel1 vs Travel3 | 7.9%       | 92.1%      |
| Travel2 vs Travel3 | 9.1%       | 90.9%      |
| Eshop1 vs Eshop2   | 7.1%       | 92.9%      |
| Eshop1 vs Eshop3   | 3.5%       | 96.5%      |
| Eshop2 vs Eshop3   | 5.8%       | 94.2%      |

*Travel Website.* If we ignore the independence assumption, a significant (p<0.01) and strong correlation was found among the distance matrices produced (Table 2). Mantel tests using 10000 permutations also found a significant (p<0.001) and strong correlation in all cases (Table 3).

*Eshop.* The distance matrices produced by the three open card sorts were significantly (p<0.01) and strongly associated (Table 4). Mantel tests using 10000 permutations also found a significant (p<0.001) and strong correlation in all cases (Table 5).

These findings provide evidence that highly similar groupings were produced by similar profile participants performing card sorting exercises for the same content.

### RQ2: Quantitative Analysis of Data from Different Card Sorting Studies

Deriving information structures from card sorting data can be a challenge. Various quantitative analysis techniques have been proposed in the literature, such as factor analysis, k-means and multidimensional scaling [1, 9, 11]. However, hierarchical cluster analysis remains the most widely used technique for this purpose [12]. Typically, the average-linkage method is employed as it usually produces balanced clusters that are easier to interpret [20]. The main output of this quantitative analysis is a tree diagram, known as dendrogram, a branching diagram representing a hierarchy of categories based on content items similarity. This dendrogram can provide insight on organization of content and menu structures.

We employed the method reported in [4, 18] to objectively compare the dendrograms produced for each set of studies. To this end, we used the amount of separation between base-clusters, i.e., the first-level clusters that are based on cards that are most similar. This is achieved as in the following: for the two cards of each base-cluster in the one dendrogram one counts the number of nodes (i.e. intersections in the dendrogram) that exist between these two cards in the other dendrogram. Then, this value is normalized against its maximum possible value, which corresponds to having to traverse all nodes in the dendrogram to connect the cards that form the base-cluster.

*Travel Website.* The total number of base-clusters was similar in all three studies: 12 for Travel1, and 13 for Travel2 and Travel3. The separation of base-clusters ranged from 4.7% to 9.1% (Table 6). In other words, the base-clusters were highly similar (from 90.9% to 95.3%).

*Eshop.* In all three studies concerning the eshop redesign, a similar number of base-clusters was formed: 17 for Eshop1, 16 for Eshop2 and 15 for Eshop3. The separation of base-clusters ranged from 3.5% to 7.1% (Table 6), which corresponds to similarity from 92.9% to 96.5%.

These findings provide evidence that the dendrograms produced by different card sorts of the same content were highly similar (from 90.9% to 96.5%) in terms of their base-clusters.

### RQ3: Navigation Schemes from Different Card Sorting Studies

One important challenge that arises in quantitative analysis of card sort data is where to cut the line in the dendrogram. This decision greatly affects the final navigation scheme. Given the same dendrogram, two people can produce different content structures. We employed a method reported in [5] to objectively compare navigation schemes that might be derived from a dendrogram.

Table 7: Elbow-based Navigation Schemes Comparisons

|  | Agreement |
|---|---|
| Travel1 vs Travel2 | 76.3% |
| Travel1 vs Travel3 | 63.2% |
| Travel2 vs Travel3 | 71.1% |
| Eshop1 vs Eshop2 | 70.9% |
| Eshop1 vs Eshop3 | 76.4% |
| Eshop2 vs Eshop3 | 92.7% |

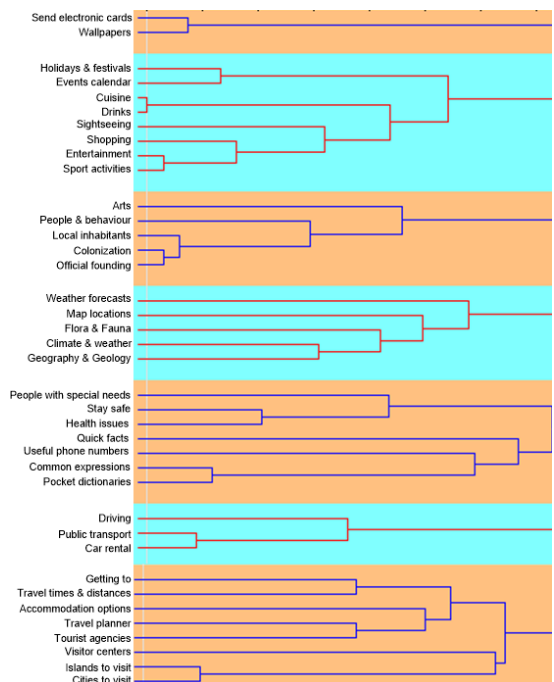

Figure 1: The elbow-based average linkage dendrogram for the Travel1 study. EZCalc [2] was used to produce the dendrogram.

This method first finds the elbow-based navigation scheme of each dendrogram, i.e., the tree structure produced by applying the Elbow criterion to find the optimal, in terms of variance explained number of clusters. Then, it calculates the agreement of these schemes operationalized as the percentage of items that were grouped together in the same category for both structures. For our study, we found the elbow-based navigation scheme for each card sort exercise and then compared the schemes proposed for the same domain (travel, eshop). As an example, Fig.1 and Fig.2 present the elbow-based navigation schemes for the Travel1 and Travel2 studies.

*Travel Website.* The elbow-based navigation schemes had a similar number of clusters; 7 for Travel1, and 8 for Travel2 and Travel3 respectively. Their mean cluster size, that is the mean number of content items grouped per cluster, was also comparable; 5.4, 4.2 and 4.2 respectively. The elbow-based navigation schemes agreement ranged from 63.2% to 76.3% (Table 7).

*Eshop.* In the three dendrograms for the eshop redesign, a similar number of clusters was calculated for the elbow-based navigation scheme; 9, 7, and 8 for the Eshop1, Eshop2 and Eshop3 respectively. The mean cluster size was also similar; 6.1, 7.9 and 6.1 respectively. The elbow-based navigation schemes agreement ranged from 70.9% to 92.7% (Table 7).

These results suggest that the elbow-based navigation schemes in each set of studies were rather similar (from 63.2% to 92.7%).

## Limitations

This research has limitations. First, analysis does not take into account qualitative data, such as users' labels for the groups or comments during the card sort. In addition, the final navigation scheme should also take into account other factors [17] such as whether one can provide labels with adequate information scent [14, 19] to the groups, and visual design constraints. These factors were not considered in this paper. Finally, additional similar studies with a varying number of cards and website domains are required to ensure the generalizability of this study's findings.

## CONCLUSIONS

In this paper, an empirical evaluation of the open card sorting's cross-study reliability is reported. To this end, data from six card sort studies were compared.

It was found that different groups of similar profile participants provided groupings for the same content items that were significantly ($p<0.001$) and strongly associated; r from 0.609 to 0.930. Quantitative analysis of card sort data using the popular hierarchical cluster analysis method was also employed. Results showed that the base-clusters formed, that is the clusters containing the most similar cards, were also highly similar; from 90.9% to 96.5%. In addition, the derived elbow-based navigation schemes were compared. These structures are obtained by cutting the dendrogram based on the optimal, in terms of variance, number of clusters (Elbow criterion). It was found that the elbow-based navigation schemes had rather high agreement; from 63.2% to 92.7%.

These findings tend to provide support for the cross-study reliability of card sorting. This is quite valuable because it justifies why a practitioner may only run one study instead of multiple.

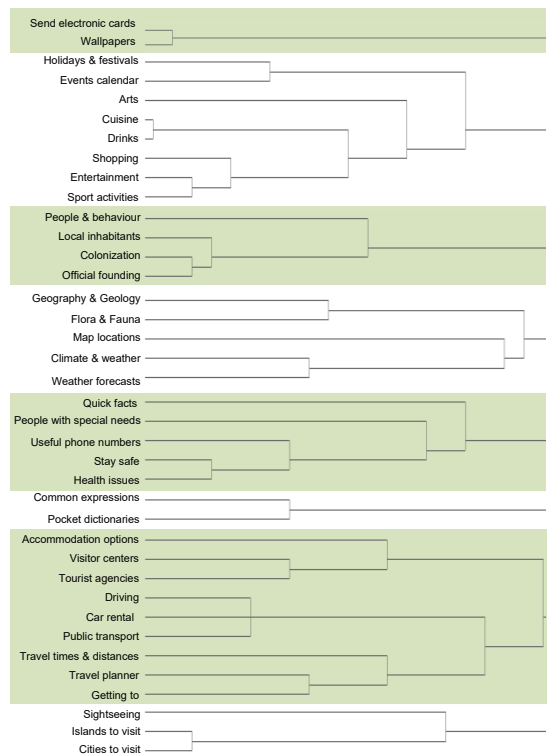

**Figure 2: The elbow-based average linkage dendrogram for the Travel2 study. OptimalSort was used to produce the dendrogram.**

It also suggests that the concern for multiple mental models [12] may not be that strong, at least for similar profile participants.

Future work includes investigation of the test-retest reliability of open card sort groupings. To this end, we will involve the same users sorting the same content items at different time intervals and compare the obtained results. Finally, we plan to investigate the real word impact of different navigation schemes on actual user experience. To this end, we will conduct user testing on existing website IAs and those produced by card sorts for the same content and then compare traditional usability measures, such as task success, time on task and perceived usability ratings.

## REFERENCES

[1] Miranda G. Capra. 2005. Factor analysis of card sort data: an alternative to hierarchical cluster analysis. In *Proc. of HFES*, 691–695.

[2] Jianming Dong, Shirley Martin, and Paul Waldo. 2001. A user input and analysis tool for information architecture. In *Ext. Abstracts of CHI 2001*, ACM Press, 23–24.

[3] James Kalbach. 2007. *Designing Web navigation: Optimizing the user experience*. O'Reilly Media.

[4] Christos Katsanos, Nikolaos Tselios, and Nikolaos Avouris. 2008. Automated semantic elaboration of web site information architecture. *Interacting with Computers 20*, 6, 535–544.

[5] Christos Katsanos, Nikolaos Tselios, and Nikolaos Avouris. 2008. AutoCardSorter: designing the information architecture of a web site using latent semantic analysis. In *Proc. of CHI 2008*, ACM Press, 875–878.

[6] N. Mantel. 1967. The detection of disease clustering and a generalized regression approach. *Cancer Research 27*, 2, 209–220.

[7] Peter Morville. 2005. *Ambient findability: What we find changes who we become*. O'Reilly Media.

[8] Peter Morville and Louis Rosenfeld. 2006. *Information Architecture for the World Wide Web*. O'Reilly Media.

[9] Ather Nawaz. 2012. A comparison of card-sorting analysis methods. In *Proc. of APCHI 2012*, ACM Press, 583–592.

[10] Jacob Nielsen. 2004. *Card Sorting: How many users to test*. Alertbox: http://www.useit.com/alertbox/20040719.html.

[11] Sione Paea and Ross Baird. 2018. Information Architecture (IA): Using multidimensional scaling (MDS) and K-Means clustering algorithm for analysis of card sorting data. *Journal of Usability Studies 13*, 3, 138–157.

[12] Celeste Lyn Paul. 2014. Analyzing card-sorting data using graph visualization. *Journal of Usability Studies 9*, 3, 87–104.

[13] Helen Petrie, Christopher Power, Paul Cairns, and Cagla Seneler. 2011. Using card sorts for understanding website information architectures: technological, methodological and cultural issues. In *Proc. of INTERACT 2011*, 309–322.

[14] Peter Pirolli. 2007. *Information Foraging theory: Adaptive interaction with information*. Oxford University Press, USA.

[15] Carol Righi, Janice James, Michael Beasley, Donald L. Day, Jean E. Fox, Jennifer Gieber, Chris Howe and Laconya Ruby. 2013. Card sort analysis best practices. *Journal of Usability Studies 8*, 3, 69–89.

[16] Per Rovegard, Lefteris Angelis, and Claes Wohlin. 2008. An empirical study on views of importance of change impact analysis issues. *IEEE Transactions on Software Engineering 34*, 4, 516–530.

[17] Donna Spencer. 2009. *Card sorting: designing usable categories*. Rosenfeld Media, Brooklyn, N.Y.

[18] Thomas Tullis and Larry Wood. 2005. How many users are enough for a card-sorting study? Presented at *UPA 2004*.

[19] Nikolaos Tselios, Christos Katsanos, and Nikolaos Avouris. 2009. Investigating the effect of hyperlink information scent on users' interaction with a web site. In *Proc. of INTERACT 2009*, 138–142.

[20] Ian H. Witten and Elbe Frank. 2005. *Data mining: practical machine learning tools and techniques*. Morgan Kaufmann., Burlington, MA, USA.

[21] Jed Wood and Larry Wood. 2008. Card sorting: current practices and beyond. *Journal of Usability Studies 4*, 1, 1–6.